\documentclass[twocolumn,preprintnumbers,amsmath,amssymb,superscriptaddress,prb]{revtex4}

\usepackage{graphicx}
\usepackage{epsfig}
\usepackage{dcolumn}
\usepackage{bm}

\begin{document}


\title{Thermal depinning of fluxons in discrete Josephson rings}

\author{J. J. Mazo}

\affiliation{Dpto. de F\'{\i}sica de la Materia Condensada,
Universidad de Zaragoza, 50009 Zaragoza, Spain}

\affiliation{Instituto de Ciencia de Materiales de Arag\'on,
C.S.I.C.-Universidad de Zaragoza, 50009 Zaragoza, Spain}

\author{F. Naranjo}

\affiliation{Dpto. de F\'{\i}sica de la Materia Condensada,
Universidad de Zaragoza, 50009 Zaragoza, Spain}

\affiliation{Instituto de Ciencia de Materiales de Arag\'on,
C.S.I.C.-Universidad de Zaragoza, 50009 Zaragoza, Spain}

\affiliation{Universidad Pedag\'ogica y Tecnol\'ogica de
Colombia, Tunja, Colombia}

\author{K. Segall}

\affiliation{Department of Physics and Astronomy, Colgate University,
Hamilton NY 13346}

\author{}

\date{\today}

\begin{abstract}
We study the thermal depinning of single fluxons in rings made of
Josephson junctions. Due to thermal fluctuations a fluxon can be
excited from its energy minima and move through the array, causing a
voltage across each junction.  We find that for the initial depinning,
the fluxon behaves as a single particle and follows a Kramers-type
escape law. However, under some conditions this single particle
description breaks down.  At low values of the discreteness parameter
and low values of the damping, the depinning rate is larger than the
single particle result would suggest.  In addition, for
some values of the parameters the fluxon can undergo
low-voltage diffusion before switching to the high-voltage whirling
mode.  This type of diffusion is similar to phase diffusion in a
single junction, but occurs without frequency-dependent damping.  We
study the switching to the whirling state as well.
\end{abstract}

\maketitle

\section{Introduction}

In past years many works have been devoted to the study of extended
discrete nonlinear systems. On the one hand, it is important to deepen
our knowledge of general properties of such systems since they often
have application to many different physical situations. On the other
hand, many physical systems are well described by nonlinear discrete
models. In this field, the emergence of the concept of soliton for
instance (in the continuous and its discrete counterparts) was
paradigmatic.~\cite{Rem94,Sco03} A well known example of model system
supporting this type of nonlinear excitations is the discrete
Sine-Gordon equation (also called Frenkel-Kontorova
model).~\cite{BK98,BK04,Flo96}

A Josephson-junction (JJ) array is by construction a discrete system
made of interacting nonlinear solid state devices. JJ arrays are an
example of physical systems with great fundamental and technological
interest which are well described by discrete nonlinear
models.~\cite{Lik86,Tin96} From the experimental point of view,
Josephson junctions are a privileged place to study solitons and to
explore their possible applications.~\cite{Ust98}

\begin{figure}[b]
\centering{\includegraphics[width=8.5cm]{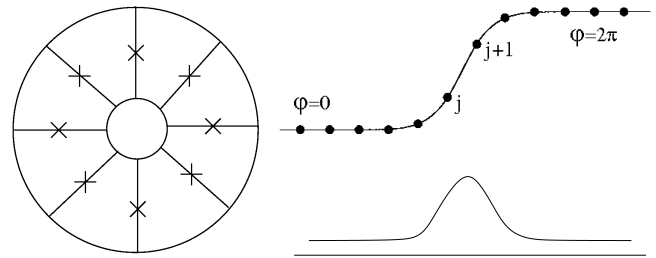}}
\caption{Left: Scheme of the JJ ring. Lines are for superconducting
wires and crosses for Josephson junctions. Right: Phase configuration
profile (top) and magnetic flux (bottom) for a fluxon in JJ ring.}
\label{fig:ring}
\end{figure}

Of the many different geometries for a JJ array the so-called
Josephson ring (a set of JJ connected in parallel and closed forming a
ring, see Fig.~\ref{fig:ring}) is well described by a discrete
sine-Gordon equation and supports nonlinear discrete solitons or
kinks, usually called fluxons in this context.~\cite{Wat95b} Thus,
many studies of the discrete sine-Gordon equation or the role of kinks
in nonlinear arrays have direct application when studying JJ
rings. Conversely the study and modelization of JJ rings allows
exploration of new issues concerning the behavior of these nonlinear
phenomena.

Our model system is the so-called Josephson ring, a collection of JJ
coupled in parallel and forming a ring (see
figure~\ref{fig:ring}). When cooled below the superconducting critical
temperature an integer number of magnetic flux quanta, fluxons, can be
trapped in the ring. Then, the physical properties of the array are
dominated by the presence of the fluxons in the system. The I-V curve
of the array shows the mean voltage across the array as a function of
a constant external current applied to every junction of the
array. When current is applied, the array remains superconducting up
to a certain critical value which defines the critical current of the
array. In the presence of fluxons this current corresponds to the
fluxon depinning current. At this current, the fluxon starts to move
around the array and finite voltage is measured.

In many cases the energy exchange between the system and the
environment is relevant, so thermal fluctuations have to be
considered.~\cite{Tin96} Because of this, at finite temperature the
value of the depinning current and the shape of the I-V curve can be
strongly affected by thermal fluctuations. Motivated by recent
experiments on the thermal depinning and dynamics of fluxons (kinks)
in small rings made of 9 junctions,~\cite{Ken08} in this work we
numerically explore some of these issues. In addition, in some cases
the fluxons can be understood as particles on a substrate periodic
potential.

The main object of this paper is to study numerically the thermal
depinning of a single fluxon in a JJ ring and compare it with
numerical simulations and analytical predictions for the case of a
single particle. We have found excellent agreement in many
cases. However, under some conditions the single particle description
fails.  In addition, for some values of the parameters the fluxon can
undergo low-voltage diffusion before switching to the high-voltage
whirling mode.  This type of diffusion is similar to phase diffusion
in a single underdamped junction, but occurs without
frequency-dependent damping.

\begin{figure*}[]
\centering{\includegraphics[width=17cm]{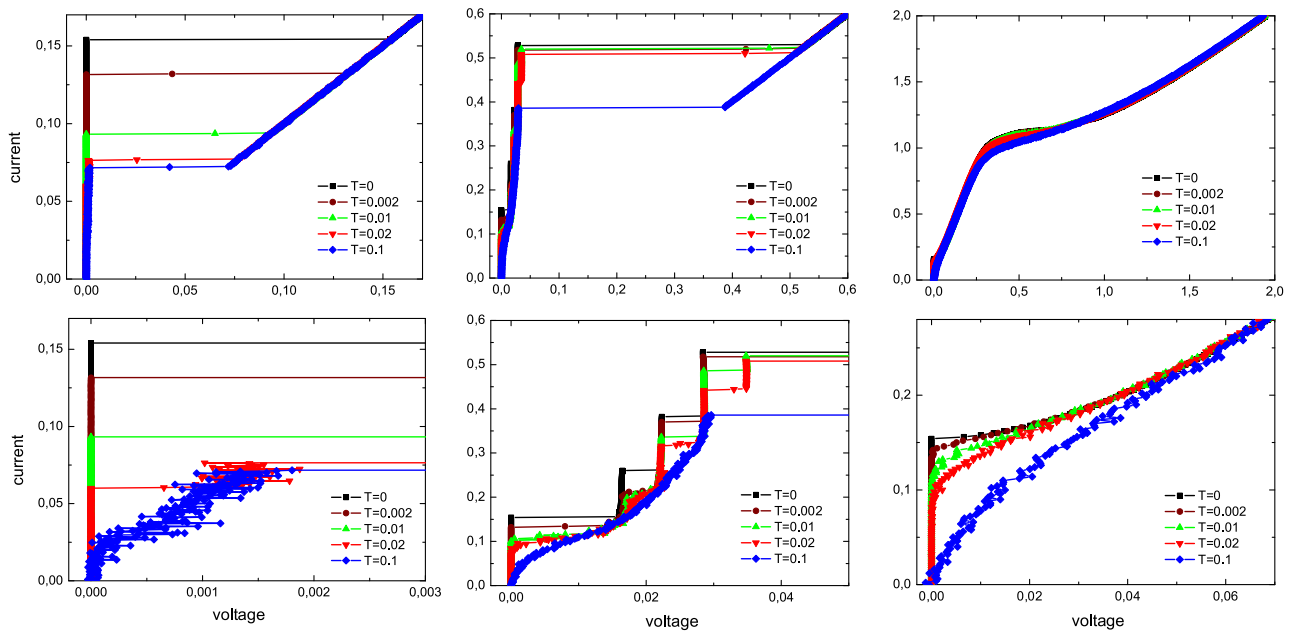}}
\caption{(Color online) $I-V$ curves for one fluxon in a 9 junctions ring with
  $\lambda=0.4$ at $\Gamma=0.01$ (left), $\Gamma=0.1$ (middle) and
  $\Gamma=1.0$ (right). Each figure shows 5 different temperatures
  ($T=0,0.002,0.01,0.02,0.1$).}
\label{fig:ivs}
\end{figure*}

\section{Equations}

Josephson junctions are made of two superconducting materials
separated by a thin insulating barrier. Driven by an external current,
this system behaves as a solid-state nonlinear oscillator and is
modeled by the same dynamical equations that describe a driven
pendulum:~\cite{Tin96} $i=\ddot{\varphi}+\Gamma
\dot{\varphi}+\sin{\varphi}+\xi(\tau)$. Here $\varphi$, the variable
that describes the behavior of the junction, is the gauge-invariant
phase difference of the superconducting order parameter at both sides
of the junction. In this equation current is normalized by the
junction critical current $I_c$ and time by the junction plasma
frequency $\omega_p=\sqrt{2\pi I_c/ \Phi_0 C}$ ($\Phi_0=h/2e$ is the
magnetic flux quantum and $C$ the junction capacitance). $\Gamma$ is
an important parameter which measures the dissipation in the system
($\Gamma=\sqrt{\Phi_0/2\pi I_c C R^2}$, with $R$ the effective
resistance of the junction). The last term, $\xi(\tau)$, describes the
effect of thermal noise in the dynamics (Johnson current noise) and
satisfies $\langle \xi(\tau) \rangle=0$ and $\langle \xi(\tau)
\xi(\tau')\rangle=2\Gamma T \delta(\tau-\tau')$ where we use $T$ for a
normalized temperature $T=k_B T_{\rm exp} / E_J$ (with $E_J$ the
Josephson energy $E_J=\Phi_0 I_C / 2\pi)$.  The normalized dc voltage
$v$ which gives the response of the system to the external current is
defined by $v=V_{dc}/I_c R=(\Phi_0/2\pi I_c R) \langle d\varphi/dt
\rangle= \Gamma \langle d\varphi/d\tau \rangle$.

As previously stated, the JJ ring consists of a series of individual junctions
connected in parallel. Such system can be though of as a series of
coupled pendula. Following the usual model for the system, the
equations for an array made of $N$ coupled junctions driven the same
external current are given by:~\cite{Wat95b}
\begin{equation}
\ddot{\varphi_j}+\Gamma \dot{\varphi_j}+ \sin{\varphi_j}+\xi_j(\tau) =
\lambda(\varphi_{j+1}-2\varphi_j+\varphi_{j-1})+i
\end{equation}
Index $j$ denotes different junctions and run from $1$ to $N$. The new
parameter $\lambda$ accounts for the coupling between the junctions
which, in the framework of this model, occurs between neighbors and
has an inductive character $\lambda=\Phi_0 / 2 \pi I_c L$, with $L$
the self inductance of every cell in the array. Boundary conditions
are defined by the topology of the array (here we consider circular
arrays) and the number $M$ of trapped fluxons in the system:
$\varphi_{j+N} = \varphi_j + 2\pi M$.

In this article we will consider the case of a single fluxon. We have
studied different sizes for the array, but here we will present
results for an array made of 9 junctions, similar to those being
experimentally studied. We will also study different values of
$\lambda$ and restrict our interest to underdamped arrays biased by a
dc current in a broad range of temperatures.

\section{Results}

In this section we are going to present numerical simulations of the
dynamics of one fluxon in a Josephson ring and one particle in a
periodic potential. We will also show numerical calculations from
single particle thermal escape theory.

\subsection{I-V curves: damping regimes}

Figure~\ref{fig:ivs} shows single I-V curves for one fluxon in a 9
junctions Josephson ring with $\lambda=0.4$. In this case the width of
the fluxon is close to 2, so it is well localized in the array and
discreteness effects are important.~\cite{disc} Curves were simulated
at three different values of damping and five temperatures. Current
was increased from zero to some maximum at an average ramp equal to
$\frac{8}{3} \times 10^{-7}$ in normalized units.

Let us look first at the $T=0$ curves. If we start at zero bias, in
all the cases the fluxon is pinned to the array up to the so-called
depinning current $i_{\rm dep}^0$ is reached (for $\lambda=0.4$,
$i_{\rm dep}^0 \simeq 0.155$). Above this current very different I-V
characteristics are observed depending on the value of the damping.

At small damping the system switches from the $v=0$ state to a
high-voltage ohmic state ($v=i$) where all the junctions rotate
uniformly (whirling branch). The damping is so small that when the
fluxon moves through the array it excites all the junctions to the
high-voltage state. In this voltage state the fluxon is totally
delocalized in the array. For clarity, we have shown only curves for
increasing current. If current is decreased from the high-voltage
state the junctions retrap at a small value of the current (mostly
defined by $\Gamma$). The curve is hysteretic and shows bistability
for a wide range of currents.

\begin{figure}[t]
\centering{\includegraphics[width=8cm]{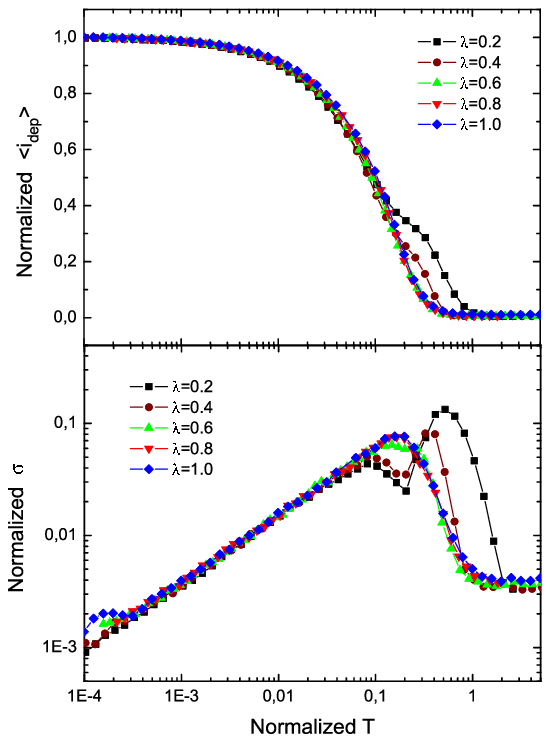}}
\caption{(Color online) Numerical calculation of normalized $\langle
  i_{\rm dep} \rangle$ and $\sigma$ versus normalized $T$ at
  $\Gamma=0.01$ and 5 different values of $\lambda$ ($\lambda=0.2,
  0.4,0.6,0.8,1.0$).}
\label{fig:g0.01}
\end{figure}

At intermediate values of damping the dynamics is much more
complex. Now the I-V curve shows a low-voltage region dominated by a
series of steps which correspond to resonances between the fluxon
velocity and the linear modes of the array. These resonances has been
the object of great attention in the
past.~\cite{Ust93,Van95,Ust95,Wat95,Wat95b,Zhe98,Pfe08} For these
currents the fluxon moves around the ring in a localized manner. This
regime persists up to a given value of the current for which the
fluxon reaches a high velocity and all the junctions switch to the
high-voltage part of the curve. At moderate damping the IV curve can
also be multistable with hysteresis loops on every step.

At high values of the damping dissipation governs the dynamics,
multistability disappears and the voltage increases from zero without
discontinuities and jumps as soon as current reaches the depinning
value. In this part of the curve a localized fluxon is moving around
the ring and voltage is related with the fluxon velocity. For currents
close to 1 (in normalized units) it starts the transient to another
regime where the fluxon delocalizes and all the junctions rotate and
contribute to the overall voltage in the array.~\cite{Flo96}

Figure~\ref{fig:ivs} also shows the dynamics of the array in the
presence of thermal noise. As can be seen in the figure, the first
thermal effect is that the fluxon depins at smaller currents. For
small damping we find that if temperature is high enough, noise also
induces a fluxon diffusion branch, which we discuss later in the
paper. At moderate damping, the low-voltage resonances are rounded and
at high enough temperature voltage increases smoothly from zero to
some value on the fluxon diffusion branch, and then switches to the
high-voltage region of the I-V curve. At high damping temperature
causes a rounding of the curve.

We will study how temperature affects the I-V curve at low damping
since this is the case for the experimental system we are trying to
model. Usually the depinning current is experimentally defined as the
current for which measured voltage is above a certain threshold. We
have followed the same definition in our simulations. This is a good
definition in the low damping and low temperature regime where the
system switches between two very different voltage values so a
threshold independent current is expected. However, the election of
the threshold voltage is not trivial. A small threshold can give
problems at large temperatures where voltage fluctuations are also
large. A large threshold is not a good choice since ignore possible
low-voltage states like the fluxon diffusion one. If low-voltage
states are present we distinguish between the depinning current
$i_{\rm dep}$ and the switching current $i_{\rm sw}$, where depinning
marks the end of the superconducting state and switching the
transition to the high-voltage branch. At very high temperatures noise
can reach the threshold level and first switching is suppressed. To
study this issue we have used in our simulation three or five
different thresholds and compare results for all of them.

\subsection{$\langle i_{\rm dep}(T) \rangle$}

The depinnig current is a stochastic variable with a given probability
distribution. We present results for the mean value of the depinning
current $\langle i_{\rm dep} \rangle$ and its standard deviation
$\sigma$. Results were obtained after the numerical simulation of the
dynamical equations of the system for 1000 samples.  We used different
values of damping $\Gamma$ (typically from 0.001 to 0.1) and coupling
$\lambda$ (usually from 0.2 to 1.0). The number of junctions in the
array is $N=9$.  Another important parameter of the simulations is the
current ramp; in our case this ramp change from one simulation to
other but it is of the order of $10^{-7}$.

Figure~\ref{fig:g0.01} shows results for $\Gamma=0.01$ and five
different values of the coupling ($\lambda=0.2, 0.4, 0.6, 0.8$ and
$1.0$). The main physical properties of the fluxon in the array change
importantly with the value of $\lambda$. Thus, the zero temperature
depinning current of the array $i_{\rm dep}^0$ decreases a factor of
30 from $\lambda=0.2$ to $\lambda=1.0$ (see table below). In order to
compare the five curves, in figure~\ref{fig:g0.01} both axis have been
scaled by the value of the zero temperature depinning current for
every case.~\cite{comment1} We see that once scaled all curves are
similar showing that in this range of parameters these results can be
understood in a unified manner.

In the figure we see that $\langle i_{\rm dep} \rangle$ decreases to
zero as effect of temperature. All the curves follow the same behavior
but at high temperatures the small $\lambda$ curves slightly deviates
from the others. For temperatures of the order of the barrier thermal
fluctuations dominate the dynamics and the depinning current goes to
zero. In fact, at high temperatures there is no a good definition of
depinning current since different thresholds may give different
results. With respect to the standard deviation, we can see that it
grows with $T^{2/3}$ as predicted by standard thermal activation
theory and reaches a maximum at high temperatures, when $\langle
i_{\rm dep} \rangle$ has an inflection point, close to $\langle i_{\rm
  dep} \rangle \to 0$. Then the standard deviation decreases since all
escape events happens in a narrow range of current values.

\subsection{The fluxon as a single particle.}

When studying the dynamics of one fluxon in the array it is very
common to use the picture of this extended and collective object as a
single particle.~\cite{BK98,BK04,Mar97,Cat01} This approach has been
extensively used in the past and, as we will see, it is very useful
although not exact.

Let us consider a new variable representing the center of masses of
the fluxon or the position of the fluxon in the array. Then, in the
simplest approach, the dynamics of a fluxon in a ring can be approach
by the dynamics of a driven, damped, massive particle experiencing a
sinusoidal substrate potential (Peierls-Nabarro potential) and
subjected to thermal fluctuations:
\begin{equation}
m \ddot{X}+\Gamma m \dot{X}+i_{\rm dep}^0 \sin{X}=i+\xi(\tau)
\label{1p}
\end{equation}
where
\begin{equation}
\langle \xi(\tau) \rangle=0 \; \; {\rm and} \; \;
\langle \xi(\tau) \xi(\tau') \rangle =
2 m \Gamma T \delta(\tau-\tau')
\end{equation}

In this simple approach we are neglecting for instance the spatial
dependence of the mass, effective damping due to the other degrees of
freedom of the system and higher order terms in the expansion of the
substrate potential for the fluxon.

Table~\ref{table} gives a relation of numerically computed values of
some of the parameters of the fluxon and its effective potential in
the single particle picture: $E_{\rm PN}$ is the zero current
potential barrier; $\omega_{\rm PN}^2$ is the squared frequency for
small amplitude oscillations of the fluxon around equilibrium; $m$ is
the fluxon effective mass at rest (computed as $m=E_{\rm
  PN}/2\omega_{\rm PN}^2$) and $i_{\rm dep}^0$ the depinning
current. For a perfect sinusoidal potential we should get $i_{\rm
  dep}^0=E_{\rm PN}/2$. The exact results are close to it.

\begin{table}
\caption{\label{table} Fluxon parameters at different values of
$\lambda$}
\begin{ruledtabular}
\begin{tabular}{cllll}
$\lambda$ \; & \; $E_{\rm PN}$ & \; $\omega_{\rm PN}^2$ & \; $m$ & \; $i_{\rm
  dep}^0$\\ \hline
0.2 \; & 0.77842 \; & 0.77405 \; & 0.5028 \; & 0.38482 \\
0.4 \; & 0.30974 \; & 0.44775 \; & 0.3459 \; & 0.15435 \\
0.6 \; & 0.12744 \; & 0.23151 \; & 0.2757 \; & 0.06367 \\
0.8 \; & 0.05539 \; & 0.11721 \; & 0.2363 \; & 0.02767 \\
1.0 \; & 0.02550 \; & 0.06041 \; & 0.2110 \; & 0.012725
\end{tabular}
\end{ruledtabular}
\end{table}

\begin{figure}[t]
\centering{\includegraphics[width=8cm]{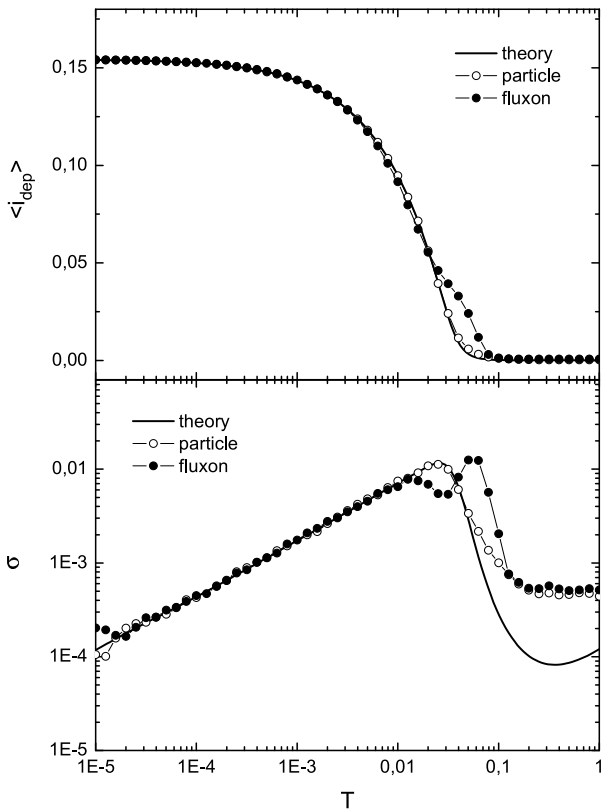}}
\caption{$\langle i_{\rm dep} \rangle$ and $\sigma$ versus $T$ at
  $\Gamma=0.01$ and $\lambda=0.4$ for the fluxon and a single particle
  in a periodic potential and comparison with the theoretical
  prediction.}
\label{fig:l04}
\end{figure}

\begin{figure}[]
\centering{\includegraphics[width=8cm]{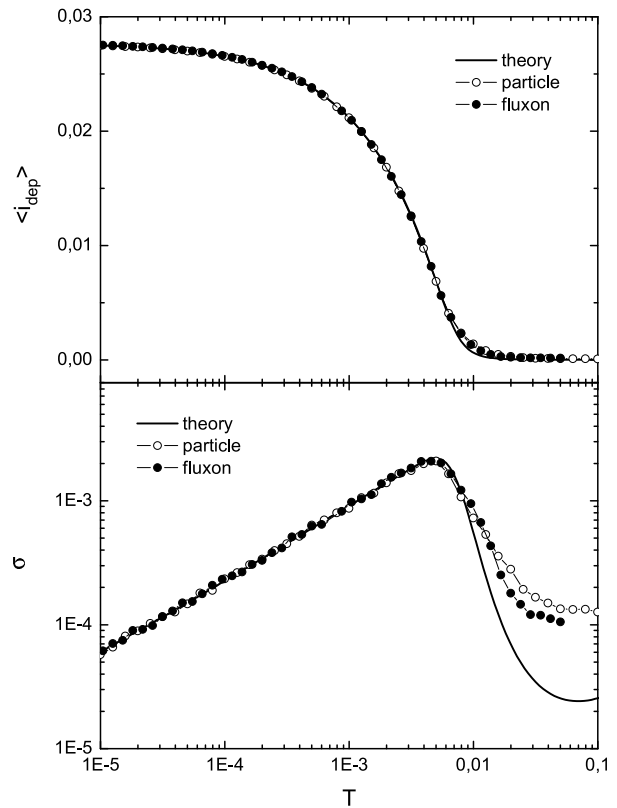}}
\caption{$\langle i_{\rm dep} \rangle$ and $\sigma$ versus $T$ at
  $\Gamma=0.01$ and $\lambda=0.8$ for the fluxon and a single particle
  in a periodic potential and comparison with the theoretical
  prediction.}
\label{fig:l08}
\end{figure}

Figures~\ref{fig:l04} and~\ref{fig:l08} show for $\lambda=0.4$ and
$\lambda=0.8$ respectively (in both cases $\Gamma=0.01$) the
comparison between the results for the fluxon in the array, numerical
simulations of the depinning of a single particle in a sinusoidal
potential (Eq.~\ref{1p}) and a theoretical calculation based on
analytical results for the thermal activation rate of particles in
sinusoidal potentials in the low damping
regimen.~\cite{Ful74,Mar87,BHL83,Mel86,Han90,Mel91} In the figure we
plot the result computed from the B\"uttiker, Harris and Landauer
equation for escape rate at $\alpha=1$ [Eq. (3.11) in
  reference~\cite{BHL83}]. We have checked that a very close result
(indistinguishable at the scale of the figure) is got when using
$\alpha=1.45\;$~\cite{1.47} or the Melnikov and Meshov
theory.~\cite{Mel86}

We can see that the single particle simulations reproduce the results
for the fluxon in an excellent way at this value of the damping in all
the temperature range although a small deviation in a range of
temperatures is observed for $\lambda=0.4$. Theoretical estimations
disagree at high values of $T$ since escape rate equations were
obtained in the infinite barrier limit ($E_b \gg k_BT$).

\begin{figure}[t]
\centering{\includegraphics[width=8cm]{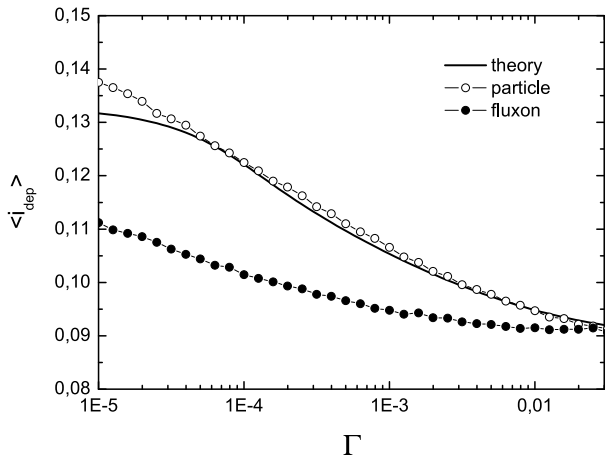}}
\caption{$\langle i_{\rm dep} \rangle$ as a function of $\Gamma$ for
$\lambda=0.4$ at $T=0.01$.}
\label{fig:cgl04}
\end{figure}

\begin{figure}[t]
\centering{\includegraphics[width=8cm]{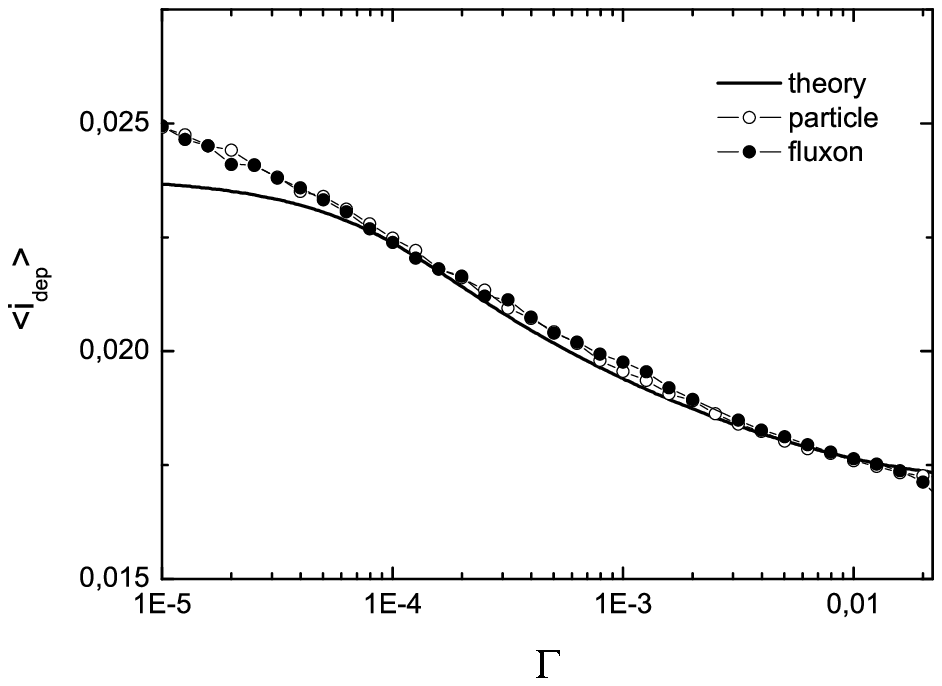}}
\caption{$\langle i_{\rm dep} \rangle$ as a function of $\Gamma$ for
$\lambda=0.8$ at $T=0.0018$.}
\label{fig:cgl08}
\end{figure}

The agreement shown in Figures~\ref{fig:l04} and~\ref{fig:l08} does
not occur for other values of coupling and damping. For instance, for
$\lambda=0.4$ and $\Gamma=0.001$ (figure~\ref{fig:g2_3} below) a
deviation of the simulated curve with respect to the theoretical
prediction is found. To study further such result we have done
numerical simulations at fixed $T$ for different values of $\Gamma$
and for $\lambda=0.4$ ($T=0.01$) and $\lambda=0.8$
($T=0.0018$).~\cite{comment3} Results are shown in
figures~\ref{fig:cgl04} and~\ref{fig:cgl08}. There we can see that for
$\lambda=0.4$ the fluxon results deviate importantly from the expected
for the single particle (or the theory) when damping is
decreased. However this is not the case for $\lambda=0.8$. In this
respect it seems to be important the degree of discreteness of the
system. Such degree is measured by the coupling parameter $\lambda$
(high $\lambda$ approach the continuum limit of the system, and small
$\lambda$ increases the discreteness effects). At small values of
$\lambda$ effects due to other degrees of freedom are more important
and if damping is small such excitations persist longer in the system.

\subsection{Fluxon diffusion}

In figure~\ref{fig:ivs} we have seen that at low damping and high
enough temperature a low-voltage branch appears in the $I-V$ curves
before the escape to the full running state.  In this low-voltage
state, the transport of the fluxon occurs through a series of
noise-induce $2\pi$ phase slips (or $2\pi /N$ depending on the
definition of the fluxon center of masses) where every jump
corresponds to a fluxon which advances one cell in the array. Such
state can not be understood in terms of the single particle
picture. In analogy with the phase diffusion that occurs in a single
junction, we label this mode of transport fluxon diffusion. In spite
of the fact of we are in the low damping regime, the fluxon is able to
travel along the ring without exciting the whirling branch. Remarkably
this is a thermally excited state and it is not seen at low
temperatures.

\begin{figure}[!t]
\centering{\includegraphics[width=8cm]{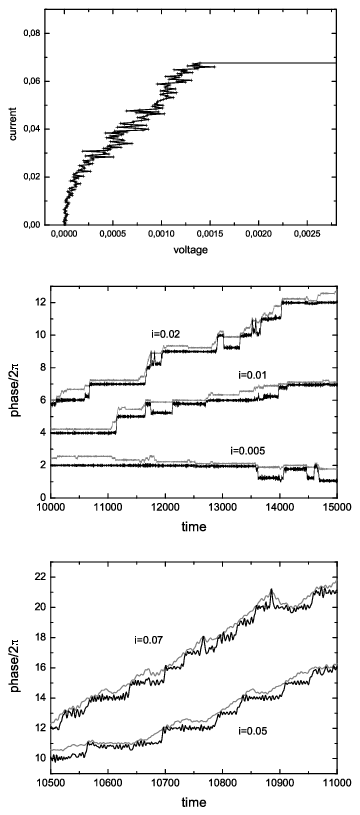}}
\caption{$\lambda=0.4$, $\Gamma=0.01$ and $T=0.1$. Top: I-V curve
  showing the small voltage fluxon diffusion branch. Medium and
  bottom: time evolution of the phase of junction 1 (black line) and
  the phase of the center of masses of the fluxon (grey line) divided
  by 2$\pi$ at $i=0.005, 0.01, 0.02, 0.05$ and $0.07$.}
\label{fig:xjt}
\end{figure}

\begin{figure}[t]
\centering{\includegraphics[width=8cm]{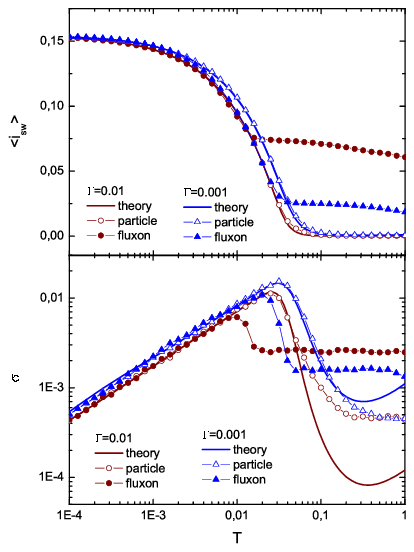}}
\caption{(Color online) $\langle i_{\rm sw} \rangle$ and $\sigma$ as a
  function of $T$ at $\lambda=0.4$, for two different values of the
  damping $\Gamma=0.01$ and $\Gamma=0.001$. We show plots for the
  theoretical prediction (solid line), the particle (open symbols) and
  the fluxon (solid symbols).}
\label{fig:g2_3}
\end{figure}

In figure~\ref{fig:xjt} we show the time evolution of the phase of one
junction (junction 1) and a phase associated with the center of mass
of the fluxon defined as $\psi= \frac{1}{N}\sum_j \varphi_j$ to allow
a better comparison. Current values have been chosen in the
low-voltage branch The random jumps of $2\pi$ for the junction or
$2\pi/N$ for the fluxon phase $\psi$ are easily observed.

We have also done numerical experiments to simulate the jump from the
diffusion branch to the full whirling state.  These simulations are
done with a higher voltage threshold, but are otherwise similar to the
previous experiments. Results for $\Gamma=0.01$ and $0.001$ are shown
in figure~\ref{fig:g2_3}, here $\lambda= 0.4$.  We see that theory and
single particle results agree quite well in all the range. At high
temperature thermal fluctuations dominate, are of the order of the
barrier, and theoretical results do not apply. We also see that the
standard deviation increases with temperature following the expected
law up to a certain value where reaches a maximum and then decreases
when $\langle i_{\rm dep} \rangle$ is close to zero.

However, in figure~\ref{fig:g2_3} we can also see that for the fluxon
curve there exists a value of $T$ such that at higher temperatures,
the value of the current for which the array switches to the whirling
branch is temperature independent. This temperature shows the
emergence of a fluxon diffusion branch in the I-V curve.  Looking at
$I-V$ curves it appears that the switching current value is defined by
some value of the fluxon velocity. Comparing the standard deviation
curve in figure~\ref{fig:g2_3} to the similar one with the lower
threshold in figure~\ref{fig:l04}, as expected, we can see that a peak
occurs much earlier in temperature for the jump from the diffusion
state.

\section{Discussion and Conclusions}

We have studied the thermal depinning of fluxons in small Josephson
rings at small values of damping. At zero temperature as current is
increased the system switches from a superconducting zero voltage
state to a resistive state where $v=i$. This happens at the so called
depinning current. Beyond this current there is not static
configuration for one fluxon in the array, and the fluxon starts to
move. Due to the low value of the damping when fluxon goes through the
array causes all the junctions to switch to a high-voltage state. Then
all the junctions do the same but with a phase difference that
accounts for the presence of one quantum of flux homogeneously
distributed along the whole array.

Due to thermal fluctuations, in an experiment the value of the
measured depinning current changes from one I-V to another and only a
probability distribution function has sense. This function is usually
characterized by its mean value and its standard deviation. The main
object of this paper has been to numerically study how these
observables behave for different system parameters (coupling
$\lambda$, damping $\Gamma$ and temperature $T$).~\cite{cr} We also
have compared these results with numerical simulations and theoretical
estimations for the depinning of a single particle in a sinusoidal
potential.

As expected, the mean value of the depinning or the switching current
decreases as temperature is raised. At low temperatures the standard
deviation follows the usual $T^{2/3}$ law. At higher temperatures the
$\sigma(T)$ function reaches a peak, that we can identify with the
points on the $\langle i_{\rm dep}(T) \rangle$ or $\langle i_{\rm sw}
(T) \rangle$ curves at which the curvature changes sign (inflection
point).

Roughly speaking our results show that the depinning of the fluxon can
be understood in terms of the stochastic dynamics of a single particle
in a tilted sinusoidal potential. However, we have seen some
unexpected effects that we attribute to discreteness. Then for the
case of small coupling ($\lambda=0.4$ and smaller) we have seen an
increasing deviation of the fluxon depinning behavior from our
expectations from the single particle picture. Arrays with a larger
coupling are closer to the continuous limit and discreteness effects
are smaller; here the single particle picture works much better.

The other effect we have observed is the emergence at small damping of
a low-voltage thermally excited state that we call fluxon
diffusion. In such cases the zero temperature I-V curve does not show
any resonance or low-voltage state and the system switches from zero
voltage to the high-voltage branch. However, after some temperature a
low-voltage branch is observed. The system first switches from zero to
this branch and then to the high-voltage state. At higher temperatures
the system first continuously increases voltage from zero (then is not
clear how to define $i_{\rm dep}$) and at larger currents switches
from the low-voltage state to the high-voltage one. We have also seen
that the values of damping and current for which this behavior is
observed depends importantly in $\lambda$ and also in the number of
junctions in the array, $N$.~\cite{N}

An important point of our work has been to compare our numerical
results with results based on the single particle picture. The main
conclusion is that for most of the cases this picture gives a good
estimation of the fluxon depinning current. In fact, in an
experimental case, where the different parameters (mainly $\lambda$,
$\Gamma$ and $I_c$) are known with some imprecision will be difficult
to identify deviations from the expected behavior. In addition, there
are some points difficult to address: the effective one-dimensional
potential for the fluxon in the array (Peierls-Nabarro potential) is
not purely sinusoidal, and the value of the fluxon mass is not
constant since it depends on the fluxon position and the current
value. We are also neglecting all the system degrees of freedom except
one and we know that in some cases this is not valid: for instance, to
understand resonant steps which are due to coupling between the fluxon
velocity and the linear waves of the discrete array or to understand
the fluxon diffusion branch. We have also considered the expression
for the escape rate in a multidimensional case. In this expression
usually the attempt frequency depends on the frequency of all the
stable modes in the minimum and the saddle.~\cite{Han90} We have
computed such numbers and check that the maximum error is smaller of
$7\%$ (and occurs for $\lambda=0.125$).

Our numerical results for the single particle agree pretty well the
predictions from Kramers theory for escape rate except for some
limits. Disagreement at high temperatures is expected since
theoretical expressions are computed in the infinite barrier limit of
the system ($E_b \gg kT$). This limit is not fulfilled at high
temperatures. This is also true at small temperatures, where most of
the escape events occur at currents very close to $i_{\rm dep}$ where
the barrier is also very small. We have checked that the $E_b/kT$
ratio in this case is also small. To finish we have to mention the
unexpected disagreement at small values of $\Gamma$. We are currently
study further such results.

In the single-particle picture, or the RCSJ model for a single
junction, it has become generally accepted that diffusion cannot
coexist with hysteresis.  This was elucidated nicely in Kautz and
Martinis~\cite{Kau90} through phase space arguments.  Simply put, if
the value of applied current is sufficient to allow a stable running
state to coexist with the fixed points of zero voltage, the basins of
attraction for that running state necessarily separates the basins of
attraction for any two neighboring fixed points.  Phase jumps between
two fixed points are thus forbidden, as the system must first pass
through the basin of attraction for the running state.  While we have
shown in previous sections that the initial escape of the fluxon from
its minima can be explained by thermal activation of a single
particle, the fluxon diffusion state in the I-V curves of
figure~\ref{fig:ivs} cannot be explained in a similar way.

In the original single-junction experiments on phase diffusion, the
coexistence of phase diffusion and hysteresis was explained by the
presence of frequency-dependent damping from the junction leads.  A
simple model of frequency-dependent damping is a series-RC circuit in
parallel with the junction, which adds an extra degree of freedom to
the phase space for the junction dynamics. This extra dimension
resolves the above-mentioned issue regarding the non-overlapping
basins of attraction. A main result of the paper is the observation of
fluxon diffusion in our simulations without frequency
dependent-damping, which has not been included in our equations.
Instead of the extra dimension introduced by frequency-dependent
damping, fluxon diffusion must occur because of the additional degrees
of freedom from the multiple junctions in the array. We plan to
explore the physics of this new diffusion mechanism in future
experiments and simulations.

When studying the behavior of the system at different temperatures we
have seen that the mean value of the distribution of the switching
current from the fluxon diffusion branch is almost constant and the
standard deviation is very small. Comparing the standard deviation
curve in figure~\ref{fig:g2_3} to the similar one with the lower
threshold in figure~\ref{fig:l04}, we also see that a peak occurs much
earlier in temperature for the jump from the diffusion state. This
peak in the standard deviation is also reminiscent of experiments on
single junctions,~\cite{Kiv05,Kras05,Man05} where a peak in the
standard deviation indicated the collapse of thermal activation and
the onset of diffusion.

To finish we want to mention that to our knowledge there are not
experimental or numerical results studying systematically the thermal
escape of fluxons or solitons in discrete arrays. However we want here
to mention the work by A. Wallraff et al on vortices in long
JJ.~\cite{Wallraff} With respect with theoretical advances we want
also to cite recent work~\cite{Sbo07} where major differences between
the macroscopic quantum tunnelling in JJ from tunnelling of a quantum
particle are reported.

\section*{Acknowledgments}

We acknowledge F. Falo for a carefully reading of the manuscript. Work
is partially supported by DGICYT project FIS2005-00337 and NSF Grant
DMR 0509450.


\begin{thebibliography}{99}
\bibitem{disc} The existence of a depinning current is also a discrete
  effect (breaking of fluxon translational symmetry). The depinning
  current rapidly approaches to zero when $\lambda$ increases and we
  approach the continuous limit of the system.
\bibitem{comment1} The physical reason for such scaling is based in
  the normalization of equation~(\ref{1p}).
\bibitem{comment3} Note that the $T/i_{\rm dep}^0$ ratio is the same
  for both cases.
\bibitem{cr} This value also depend on the current ramp which enters
  into the equations in a way well established by theory.
\bibitem{N} This paper only shows results for 9 junctions
  arrays. However, we have done also numerical simulations for larger
  arrays.
\end{thebibliography}
\end{document}